\documentclass[final,1p,times]{elsarticle}
\usepackage{color}
\usepackage[T1]{fontenc}
\usepackage[utf8]{inputenc}
\usepackage{geometry}
\usepackage{lmodern}
\usepackage{graphicx}
\usepackage[table,xcdraw]{xcolor}
\usepackage{natbib}

\journal{Astroparticle Physics}
\def\lsim{\mathrel{  %% \mathrel makes \lsim look like other relation symbols
        \raise0.3ex\hbox{$<$}\kern-0.75em{\lower0.65ex\hbox{$\sim$}}}}

\begin{document}
\begin{frontmatter}

\title{Diffusion of Cosmic Rays in a Multiphase Interstellar Medium Swept-Up by a Supernova Remnant Blast Wave}
\author[label1]{Soonyoung Roh} \ead{soonyoung@nagoya-u.jp}
\author[label1]{Shu-ichiro Inutsuka\corref{cor1}} \ead{inutsuka@nagoya-u.jp}
\author[label2]{Tsuyoshi Inoue} \ead{tsuyoshi.inoue@nao.ac.jp}
\address[label1]{Department of Physics, Graduate School of Science, Nagoya University, Nagoya 464-8602, Japan}
\address[label2]{Division of Theoretical Astronomy, National Astronomical Observatory of Japan, Osawa, Mitaka district, Tokyo 181-8588, Japan}

\begin{abstract}
Supernova remnants (SNRs) are one of the most energetic astrophysical events and 
are thought to be the dominant source of Galactic cosmic rays (CRs). 
A recent report on observations from the Fermi satellite has shown a signature 
of pion decay in the gamma-ray spectra of SNRs.
This provides strong evidence that high-energy protons are accelerated in SNRs. 
The actual gamma-ray emission from pion decay should depend on the diffusion of 
CRs in the interstellar medium. 
In order to quantitatively analyse the diffusion of high-energy CRs from 
acceleration sites, we have performed test particle numerical simulations 
of CR protons using a three-dimensional magnetohydrodynamics (MHD) simulation 
of an interstellar medium swept-up by a blast wave. 
We analyse the diffusion of CRs at a length scale of order a few pc 
in our simulated SNR, and find the diffusion of CRs is precisely described by 
a Bohm diffusion, which is required for efficient acceleration 
at least for particles with energies above 30 TeV 
for a realistic interstellar medium. 
Although we find the possibility of a superdiffusive process (travel distance 
$\propto t^{0.75}$) in our simulations, its effect on CR diffusion 
at the length scale of the turbulence in the SNR is limited. 
\end{abstract}

\end{frontmatter}

\section{Introduction}
Supernova remnants (SNRs) have long been believed to be the source of hadronic 
Galactic cosmic rays (GCRs) up to energies of the `knee', near $5 \times 10^{15}$ eV, 
of the cosmic ray (CR) spectrum. Supernova explosions forming collisionless shock waves 
induce the shocked gas and relativistic particles (hereafter cosmic rays) 
that produce multi-wavelength thermal and nonthermal emission. 
Diffusive shock acceleration (DSA) is the most promising mechanism for converting 
the kinetic energy of a supernova explosion into energetic particles 
\citep{Bell1978a, Bell1978b, Blandford1978, Axford1977} and plays an important role 
in nonthermal emission during the overall process, 
(e.g., \citep{Axford1977, Axford1982, Drury1981}). 

In the framework of DSA, an individual charged particle experiences many collisions 
with background electromagnetic waves and gains energy by shock crossing. 
This leads to a nonthermal CR spectrum of the power-law form 
$N(\varepsilon) \sim \varepsilon^{-2} $. Shock acceleration by DSA 
in SNR shocks is associated with transport processes, and some of the highest energy 
CRs eventually escape from their acceleration sites by a so-called diffusion process 
due to interactions with turbulent magnetic fields. 
To analyse the diffusion process we have to determine the effective diffusion coefficient
 $D(\textbf{r},\textbf{p})$, and the understanding of $D(\textbf{r},\textbf{p})$ is necessary to interpret many astronomical observations.

Several studies of escaping CRs have already been done and a strong spatial 
correlation between TeV emission and the molecular gas distribution 
at the Galactic Center has been observed 
\citep{Aharonian2006, Wommer2008, Ballantyne2007}. 
The pion-decay signature in SNRs is believed to be evidence for protons accelerated 
in middle-aged SNRs interacting with molecular clouds, 
\citep{Ackermann2013,Abdo2009, Abdo2010a, Abdo2010b, Abdo2010c, Tavani2010, Ohira2012}. 
Recent observations suggest that gamma-ray emission and CO+HI emission are 
spatially correlated in young SNRs \textit{RX J1713.7-3946} and 
\textit{RX J0852.0-4622} (\citep{Fukui2012,Fukui2013}, see also \citep{Abdo2011, Inoue2012-a, Gabici2014}. 

Nevertheless, the identification of pion-decay gamma rays is difficult 
because high-energy electrons also produce gamma rays via bremsstrahlung 
and Inverse Compton scattering (leptonic model) \citep{Ackermann2013}. 
X-ray observations show that electrons are accelerated to highly 
relativistic energies in SNR shocks \citep{Koyama1995}. 
Therefore, in order to understand the acceleration sites of CRs, 
it is crucial to distinguish GeV-TeV emission from Inverse Compton scattering 
by CR electrons and the decay of neutral pions produced by inelastic collisions 
between CR protons and ambient thermal nuclei.

In this paper, we investigate the diffusion of CRs using a hydrodynamics simulation 
of a strong shock wave propagating in a realistic multiphase interstellar medium and a one-phase medium. 
The organization of the paper is as follows. 
In Section 2, we describe the three-dimensional hydrodynamics simulations and the resulting configuration of electromagnetic field used.
We also briefly introduce the process of Bohm Diffusion. 
The results of test particle simulations performed in these environments 
are shown in Section 3. 
We investigate the properties of escaping CRs in terms of 
the diffusion coefficient in both energy and configuration space, 
and finally we summarize and discuss our findings in Section 4.

\section{Methods}
\subsection{Setup of Background Interstellar Medium}
\subsubsection{Multiphase medium}

The interstellar medium (ISM) is an open system in which radiative cooling 
and heating are effective. 
It is an inhomogeneous, multiphase system in which gases of different temperatures, densities, and ionization fractions can coexist in approximate pressure equilibrium.
Diffuse warm gas (diffuse intercloud gas) with T $\simeq$ 10$^4$ K and HI clouds 
(interstellar clouds) with T $\simeq$ 10$^2$ K are approximately 
in pressure equilibrium in a typical ISM environment. 
As a consequence of the thermal instability driven by external compressional events 
such as shock waves due to expanding HII regions or very late phase SNRs, 
unstable gas evolves into diffuse gas and HI clouds 
\citep{Inutsuka2005, Inoue2008-687, Inoue2012-759}. 
Therefore, inhomogeneities inevitably emerge and remain ubiquitous in the ISM. 
The characteristic length scale of an inhomogeneity can be expressed 
in terms of the ``Field length", 
which is the critical wavelength of the thermal instability \citep{Koyama2004,Field1965}. 
The Field length depends on density and temperature and can be smaller than 1 pc. 
A blast wave generated by supernova expansion sweeps up the dense and 
clumpy HI clouds of the multiphase ISM, which eventually generates 
strong velocity shear in the magnetic fields. 
Magnetic fields undergo amplification from their typical strength of $\mu$G to mG 
due to the turbulent dynamo in the post-shock region (\citep{Inoue2010,Inoue2012,Inoue2013,Giacalone2007}). 
This process may explain the existence of magnetic fields of mG strength 
investigated by Y. Uchiyama et al. (2007) \citep{Uchiyama2007}. 

T. Inoue et al. (2012) performed ideal three-dimensional magnetohydrodynamic (MHD) simulations of a strong shock wave ($v_{sh}\sim$ 2500 km s$^{-1}$) propagating in a realistic multiphase ISM as the pre-shock region. We use the data for the perpendicular shock of T. Inoue et al. (2012) at $t$ = 750 years as the background ISM to set up the electromagnetic field for our microscopic particle simulations. 

To generate a multiphase ISM, T. Inoue et al. (2012) solved the ideal MHD equations including cooling, heating, and thermal conduction, which determine the unstable scale of thermal instability. They considered a net cooling function and photoelectric heating, and generated an inhomogeneous medium via thermal instability. 
The simulation considered ideal gas and used an adiabatic index of $\Gamma$ = 5/3. The mean number density, initial thermal pressure, and initial magnetic field strength were taken to be $\langle n_0 \rangle$ = 2.0 cm$^{-3}$, $p/k_B$ = 2887 K cm$^{-3}$, and $B_{0y}$ = 5.0 $\mu$G, respectively, at the $\textit{x}$ = 0 boundary plane. 
For the density, they imposed random density fluctuations with a thermally unstable state in the range 10 K $\leq$ T $\leq$ $10^4$ K for effective cooling and heating.
In the resulting clumpy cloud, they induced a high Mach number shock wave using a hot plasma with $p_h/k_B$ = 10$^9$ K cm$^{-3}$ and $\langle n_h \rangle$ = 0.1 cm$^{-3}$. 

\subsubsection{One-phase medium}

T. Inoue et al. (2013) performed ideal three-dimensional magnetohydrodynamic (MHD) simulations 
to investigate the interaction between blast wave 
($v_{sh}\sim1800$ km $\rm s^{-1}$) and interstellar density fluctuations. 
They investigated the magnetic field amplification and the magnetic field distribution 
of turbulent SNRs driven by the Richtmyer-Meshkov instability (RMI). 
They assumed an adiabatic gas with adiabatic index $\Gamma$=5/3 and used 
a high Mach number shock wave. 
Density fluctuations superposed by sinusoidal functions were included and 
followed an isotropic power-law spectrum with random phases. 
The power spectrum of the density fluctuations was shown to be described by 
an isotropic power law for the wavenumber \textit{k} in the inertial range of turbulence:
$P(k) = \rho^2_k k^2 \propto k^{-5/3},$ 
where $\rho_k$ is the Fourier component of the density. 
The mean number density, the initial thermal pressure, and the initial magnetic field 
strength were taken to be $\langle n_0 \rangle$ = 0.5 cm$^{-3}$, 
$p/k_B$ = 4 $\times$ 10$^3$ K cm$^{-3}$, and $B_{0}$ = 3.0 $\mu $G, respectively. 
The parameters represent typical values in the diffuse ISM \citep{Myers1978, Beck2001}. 
To induce the blast wave, a hot plasma is set up as follows: 
$p_h/k_B$ = 2 $\times$ 10$^8$ K cm$^{-3}$, $\langle n_h\rangle$ = 0.05 cm$^{-3}$, 
and $B_{0y}$ = 3.0 $\mu $G at the $x$ = 0 boundary plane. 
This creates the primary shock wave whose normal vector is perpendicular to 
the mean magnetic field. The SNR is modelled as a young SNR (age = 10$^{3}$ years) 
with a late with velocity of 1800 km $\rm s^{-1}$. 
We used the data for the perpendicular shock on model 1 in \citep{Inoue2013} at $\textit{t}$ = 700 years as the background ISM.

\begin{figure}[h]
\centering
\includegraphics[scale=0.5,angle=90]{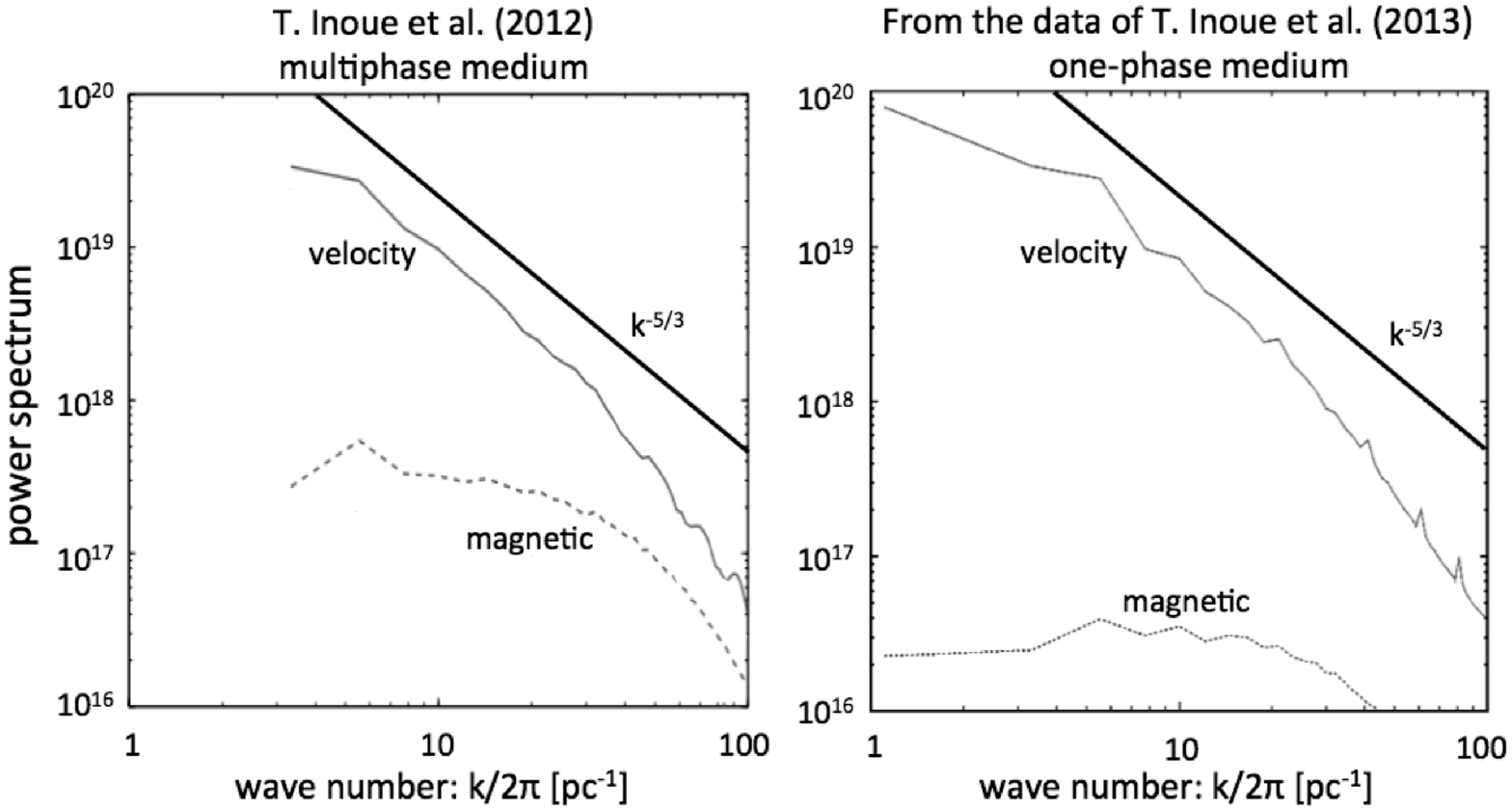}
\caption{Fourier power spectrum of SNR turbulence. Figure on left and right indicate for multiphase and one-phase medium, respectively. The upper thin solid line in figure indicates the velocity field and dashed line indicates the magnetic field. The uppermost thick solid line in figure represents the Kolmogorov law $k^{-5/3}$.}
\label{fig:FourierPowerSpectrum}
\end{figure}

A simulation box with size $L_{box}$ = 2 pc is used, and the system resolution is 
$\Delta x = L_{box}$/(number of grid cells) = $1.95 \times 10^{-3}$ pc, 
where the number of uniform grid cells is $1024^3$. 
Periodic boundary condition is used for the \textit{yz}-plane. 
The above simulation resolution and boundary condition are applied for 
both media. 
The Fourier power spectra of SNR turbulence for both media are given in 
Figure \ref{fig:FourierPowerSpectrum}. (for more details see \citep{Inoue2012, Inoue2013}.)

\subsection{Test particle simulations}
The magnetized turbulent medium flows along the (positive) \textit{x}-direction. 
We calculate the trajectories of CR particles using a snapshot of the MHD simulation data. 
We consider a collisionless environment for CRs because the mean free path, 
$\lambda = 1/n\sigma$, of relativistic particles is much larger than the gyroradius of CR particles. 
In the framework of non-relativistic ideal MHD, electromagnetic fields 
can be expressed as\\
\begin{eqnarray}
{\textit{\textbf{E}}}(t,\textit{\textbf{r}}) &=& - \frac{{\textit{\textbf{V}}(t,\textit{\textbf{r}})}_{fluid} \times {\textit{\textbf{B}}\textit{(t,\textbf{r})}}}{c},
\label{eq:E}
\end{eqnarray}
where ${\textit{\textbf{V}}}$ is the fluid velocity, ${\textit{\textbf{B}}}$ is the local magnetic field at the position of the particles, and $c$ is the speed of light. 
When we compute the Lorentz force acting on each CR particle, we interpolate ${\textit{\textbf{V}}}$ and ${\textit{\textbf{B}}}$ at the position of 
the particle before we calculate ${\textit{\textbf{E}}}$ using Equation (\ref{eq:E}). 
To trace the trajectories of CR, we solve the relativistic equation of motion using 
a fourth-order Runge-Kutta method for updating particle information. 
The momentum of a particle is defined as\\
\begin{eqnarray}
{\textit{\textbf{p}}} &=& \gamma m_{i} \textbf{\textit{v}},
\end{eqnarray}
where ${\textit{\textbf{v}}}$ is the velocity of a particle. 
$\gamma$ is the particle Lorentz factor, and $m_i$ is the mass of particle $i$. 
The equations governing the motion of relativistic charged particles are expressed as
\begin{eqnarray}
\frac{d{\textit{\textbf{p}}}}{dt} &=& \textit{q}_{i}\left( {\textit{\textbf{E}}} + \frac {{\textit{\textbf{v}}} \times {\textit{\textbf{B}}}}{c}\right), \\
\frac{d{\bf r}}{dt} &=& \frac{{\textit{\textbf{p}}}}{\gamma m_{i}},
\end{eqnarray}
where $q_i$ is the charge of particle $i$. 
The gyroradius, $\textit{R}_{g}$, of particle $i$ can be expressed by
\begin{equation}
R_{g} \approx \frac{\gamma m_{i}c^2}{|\textit{q}_i|B}~~{\rm for}~v \approx c. 
\label{eq:Rg}
\end{equation}
The time discretization, $\Delta t$, for each particle is as follows:
\begin{eqnarray}
t_{r} = \Omega^{-1}_{g}/N_{\textit{d}}
\label{eq:time}
\end{eqnarray}
\begin{eqnarray}
t_{s} = \frac{\Delta x}{N_{\textit{d}} \cdot max(|v_{x}|,|v_{y}|,|v_{z}|)}
\end{eqnarray}
\begin{eqnarray}
\Delta t = min(t_{r},t_{s}),
\end{eqnarray}
where $\Omega_{g}$ = ${|\textit{q}_i|B}/{\gamma m_{i} c} $ denotes 
the particle gyrofrequency and $\Delta \textit{x}$ is the size of a grid cell.
$N_{d}$ is the number of divisions which determines the resolution of a particle trajectory. 
The time discretization is determined 
by inverse of gyrofrequency as it shows in Equation (\ref{eq:time}). 
To trace the particle trajectory, $\Delta t$ should be smaller than gyrofrequency. 
$t_{r}$ is selected in our simulation when the particle energy is 
sufficiently small such that $R_{g} < \Delta x$, 
whereas $t_{s}$ is selected for higher energy cases satisfying $R_{g} > \Delta x$. 
For more details on $R_{g}$, see Figure \ref{fig:multi-Rg} in Section 3.2. 
We obtain $\Delta t$ by considering $ \Omega_{g} $ and the velocity components 
($v_{x}$, $v_{y}$, and $v_{z}$) of a particle as well as using 
a sufficiently large value of $N_{d}$ for tracing the trajectories of particles. 
For example, $\Delta t$ was about 250 s for $\varepsilon_0$ = $10^{3}$ TeV. 
In all calculations, we used $N_{d}$ = 10$^3$, which is the value for keeping 
the errors of Runge-Kutta time-integration sufficiently small ($< 10^{-15}$ eV) 
for energy conservation.

\section{Results of Numerical Simulations for Diffusion of Cosmic Rays}
To calculate the transport of particles, we released sets of 10$^{3}$ protons 
with the same initial energies but isotropic velocity distributions 
in the computational frame. 
The range of the initial energies ($\varepsilon_0$) considered is 
10$^{12}$ eV to 10$^{15}$ eV at intervals of $\log$ $\varepsilon_0$ = 0.5 eV. 
We assigned particle positions randomly in a cubic box region in the post-shock region (see Tables 1 and 2). 
Initially the directions of velocities are also random. 
To obtain the background fluid velocity and magnetic field we applied 
a linear interpolation 
to the MHD simulation data because the magnetic and fluid fields are defined 
at discrete points. 
We use periodic boundary conditions in the \textit{yz}-direction as was done 
in the MHD simulation. We do not follow the time integration for orbits of particles 
that have escaped the $x$-boundary. 
First, we show the results of a multiphase medium case.

\subsection{Diffusion process in a Multiphase Medium}
\subsubsection{Diffusion in Energy Space}
In a turbulent medium, second-order Fermi acceleration is expected where some particles gain energy stochastically for a sufficiently long time interval.
This can be considered to be a random walk in energy space ($\Delta \varepsilon \propto \sqrt{t}$).

\begin{figure}[h]
\centering
\includegraphics[scale=0.5]{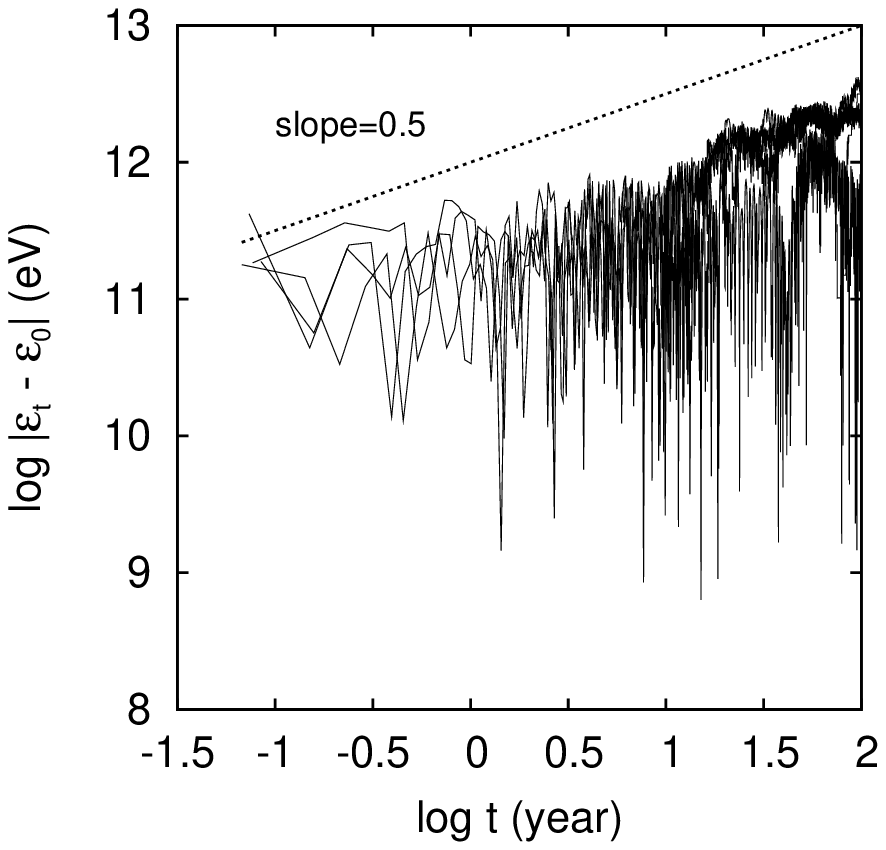}
\includegraphics[scale=0.5]{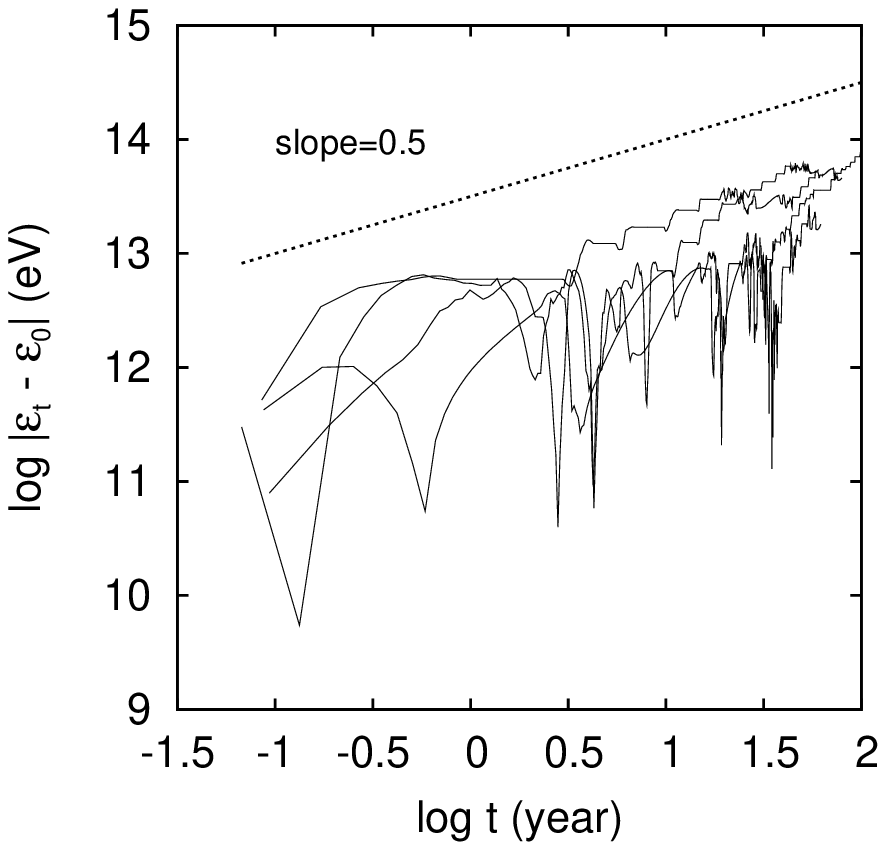}
\caption{The magnitude of the change in particle energy, $\Delta \varepsilon$, as a function of $\log t$ for protons with an initial energy of $\varepsilon_0$ = $10^2$ TeV (left) and $\varepsilon_0$ = $10^3$ TeV (right). We show several cases (solid line) for this Figure  The uppermost dotted line in figure shows the slope $t^{0.5}$.}
\label{fig:multiphase-energygain}
\end{figure}

We calculated up to $10^2$ years for $10^2$ TeV and $10^3$ TeV protons.
Figure \ref{fig:multiphase-energygain} show the magnitude of the particle energy, $\Delta \varepsilon$, as a function of $\log t$. 
Dashed lines in both figures represent energy gain proportional to $t^{0.5}$.
This demonstrates the stochastic behaviour of energy gain in our simulation.
However, the energy gain rate is very small within 10$^2$ years, less than $1\%$ of the initial energy.
Thus, we describe the energy in this paper with initial energy ($\varepsilon_0$). 

\subsubsection{Diffusion in Configuration Space}
\begin{figure}[h]
\centering
\includegraphics[scale=1]{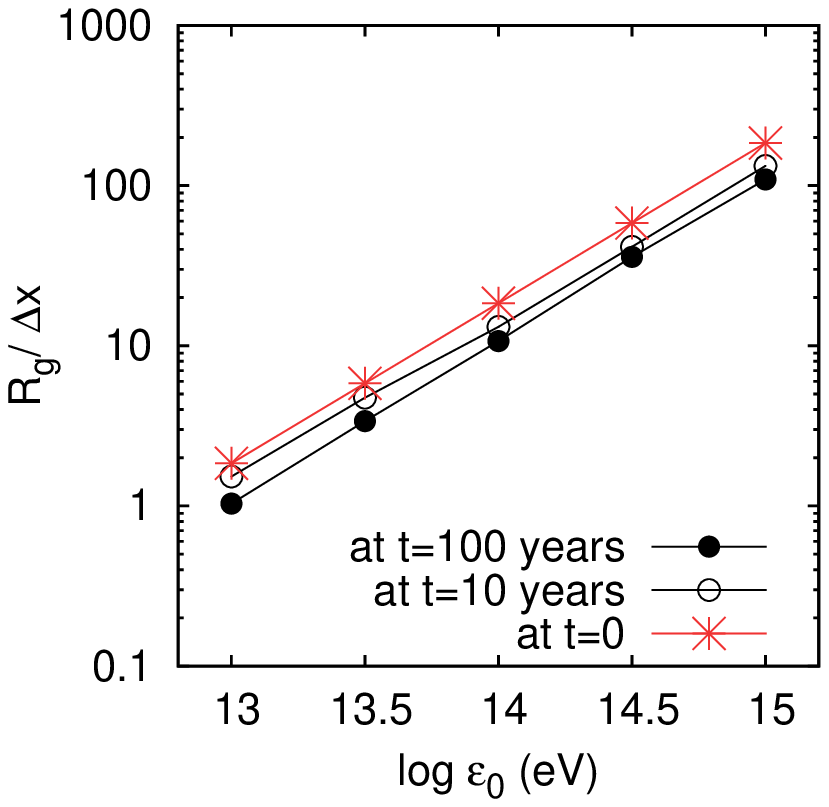}
\caption{The gyroradius ($R_g$) in units of pc as a function of cosmic ray initial energy ($\varepsilon_0$). 
The results at 10 years and 10${^2}$ years are shown by black open and filled circles, respectively. Red star symbol is for the case $t$ = 0.}
\label{fig:multi-Rg}
\end{figure}

Gyroradius (Equation (\ref{eq:Rg})) in units of pc for different values of $\varepsilon_0$ 
shows in Figure \ref{fig:multi-Rg}. 
The values are calculated for 10$^3$ particles at five different values 
of $\varepsilon_0$ at intervals of $\log \varepsilon_0$ = 0.5 eV. 
Each symbol of the solid lines corresponds to the gyroradius at a different time. 
Black open and filled circles correspond to $t$ = 10 years and 10${^2}$ years, respectively. 
Red stars indicate for an initial time. 
Note that the smallest ($L_{min}$) and largest ($L_{max}$) turbulent scales in 
our simulation correspond to a few grids and a few hundreds grids, respectively \citep{Inoue2012}. 
The grid size corresponds to $\Delta x$ = 6.03 $\times$ 10$^{15}$ cm. 
Particles are supposed to be reflected by the turbulent field 
when $L_{min} < R_{g} < L_{max}$. 
For the case of $R_g < L_{min}$, particles only see 
an almost uniform field due to grid based method, and 
thus are expected to undergo $\textbf{\textit{E}} \times \textbf{\textit{B}}$ drift. 
In this case the travel distance is proportional to $t$ at early times. 
In Figure \ref{fig:multi-Rg}, particles with $\varepsilon_0$ = 10$^{13.5}$ eV, 
10$^{14}$ eV, 10$^{14.5}$ eV, and 10$^{15}$ eV satisfy $L_{min} < R_{g} < L_{max}$. 
We do not consider the case of $\varepsilon_0$ = $10^{15}$ eV 
for our study of diffusion process in configuration space, 
because the number of escaped particles in our simulation is large.

To analyze the diffusion in configuration space
we measure the displacement 
of transported particles. 
The displacement of a CR particle is expressed by
\begin{eqnarray}
\Delta \textbf{R}_j &=& \textbf{R}_j(t) - \textbf{R}_j(t_0),
\end{eqnarray}
where the notation $j$ denotes the particle $j$ and $R=\sqrt{\bar{x}^2 + \bar{y}^2 + \bar{z}^2}$, $\bar{x}=x(t)-x(t_0)$. The dispersion of displacement R, hereafter $\sqrt{\langle(\Delta R)^2\rangle}$, is expressed by
\begin{eqnarray}
\sqrt{\langle(\Delta R)^2\rangle} &=& \sqrt{\frac {\sum\limits_{j=1}^{N_{ptls}}  (\Delta \textit{\textbf{R}}_{j})\ensuremath{^2}}{N_{ptls}}},
\end{eqnarray}
where $N_{ptls}$ denotes the total number of particles. 

\begin{figure}[h]
\centering
\includegraphics[scale=1]{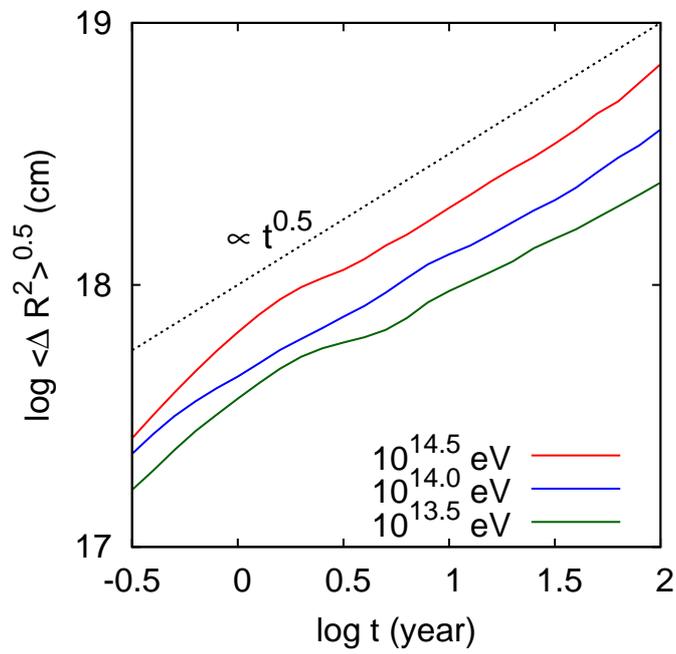}
\caption{Travel distance ($\sqrt{\langle(\Delta R)^2\rangle}$ ) of cosmic ray particles ($\varepsilon_0$) as a function of $\log \textit{t}$. 
Green, blue, and red lines indicate the curves of  
$\varepsilon_0$ = $10^{13.5}$ eV, $10^{14}$ eV and $10^{14.5}$ eV, respectively.
All lines are similar to the uppermost black dashed line that is proportional to  t$^{0.5}$, corresponding to the slowest process.
The travel distances with -0.5 < $\log \textit{t}$ < 0.5 are proportional to $t$.}
\label{fig:multi-rms-main}
\end{figure}

Here, we show the transport property of particles in 
our simulations as a function of $\log t$, and analyze the diffusion in 
configuration space. The lines in Figure \ref{fig:multi-rms-main} represent  
$\sqrt{\langle(\Delta R)^2\rangle}$ 
for protons with energies of $\varepsilon_0$ = $10^{13.5}$ eV, $10^{14}$ eV, 
and $10^{14.5}$ eV, corresponding to $R_g$ (gyroradius) $>$ $\Delta x$ (grid size) 
(see Figure \ref{fig:multi-Rg}). 
The values are averaged over $N_{ptls}$ = 10$^3$ particles for each set of 
energies ($\varepsilon_0$). We calculated up to $t$ = 10$^2$ years. 
To study the effect of changing the initial location, we defined several boxes 
for the initial CR particle locations. 
Table \ref{table:physical-multi} shows the physical values of 
the background MHD data in the specific regions, 
defined as three-dimensional boxes. 
For the study of diffusion in a multiphase medium, we describe the particle distribution and background information 
for the case of a box of size (0.3 pc)${^3}$ corresponding to region-I, as given in Table \ref{table:physical-multi}. 

The evolution of $\sqrt{\langle(\Delta R)^2\rangle}$ in Figure \ref{fig:multi-rms-main}
is described by standard diffusion, the so-called Bohm diffusion, 
that is the slowest process to be proportional to $t^{0.5}$ for all energies in our simulations, 
$\varepsilon_0$ = $10^{13.5}$ eV, $10^{14}$ eV, and $10^{14.5}$ eV. 
The black dotted line in Figure \ref{fig:multi-rms-main} is the slope $t^{0.5}$. 
For efficient acceleration of CRs, small diffusion coefficient is required.
Thus, our results can demonstrate that the diffusion process of CRs in SNRs is can be approximated by Bohm diffusion, which may support the assumptions in \citep{Inoue2012}.

\subsubsection{Comparison with Bohm Diffusion Rate}
In this Section, we quantify the diffusion coefficient of CR particles. 
According to the standard assumption of the theory of DSA, accelerated particles 
diffuse out in the plasma, where the diffusive flux obeys Fick's law:
the flux of particles is proportional to the gradient of flux density \citep{Kirk2001}. 
It is based on the assumption that charged particles are diffuse isotropically by 
turbulent electromagnetic fields.
 
In general, the diffusion coefficient can be expressed as 
\begin{eqnarray}
D(\textbf{\textit{r}},\textbf{\textit{p}}) = \frac{1}{3} \lambda \textit{\textbf{v}},
\end{eqnarray}
where $\lambda$ and $\textbf{\textit{v}}$ are defined as the mean scattering 
length and the particle velocity, respectively. The frequently used choice for 
the diffusion coefficient in DSA is that of Bohm diffusion, expressed as
\begin{eqnarray}
\lambda \approx R_{g} \left(\frac{B}{\delta B}\right)^2.
\label{eq:skilling}
\end{eqnarray}

This depends on the degree of turbulence (\citep{A.Shalchi-a,Hussein2014}). 

In our MHD simulation, however, 
the field strength shows the relation $\delta B \sim B$, as shown in Table 1.
Therefore we simply use a simplified expression for Bohm diffusion with  
$ \lambda \approx R_{g}$:
\begin{eqnarray}
\label{eq:Dbohm}
D_{Bohm}(\textbf{\textit{r}}, \textbf{\textit{p}}) = 
\frac{1}{3} R_{g} \textbf{\textit{v}} \sim \frac{1}{3} R_{g}c .
\label{eq:Dbohm}
\end{eqnarray}

This model, $D_{Bohm}(\textbf{\textit{r}},\textbf{\textit{p}})$, has been used 
by previous work (e.g., \citep{Skilling1975}). 

In order to quantify the diffusion of CR particles in our background field, 
we define $D_{sim} (t)$ as 
\begin{eqnarray}
\label{eq:Dsim}
D_{sim} (t) = \frac{\langle(\Delta R)^2\rangle}{{\textit{6t}}}.
\label{eq:Dsim}
\end{eqnarray}

We compare this quantity to the Bohm diffusion coefficient. 
If ${\langle(\Delta R)^2\rangle}$ is proportional to $t$, $D_{sim} (t)$ is constant 
in time and can be used as the diffusion coefficient in the 
standard diffusion equation for CR particles. 
Strictly speaking, $D_{sim} (t)$ is not the diffusion coefficient if it is not 
constant in time. 
Even in that case, however, $D_{sim} (t)$ is a measure of the diffusion of 
CR particles. 

\begin{figure}[h]
\centering
\includegraphics[scale=1]{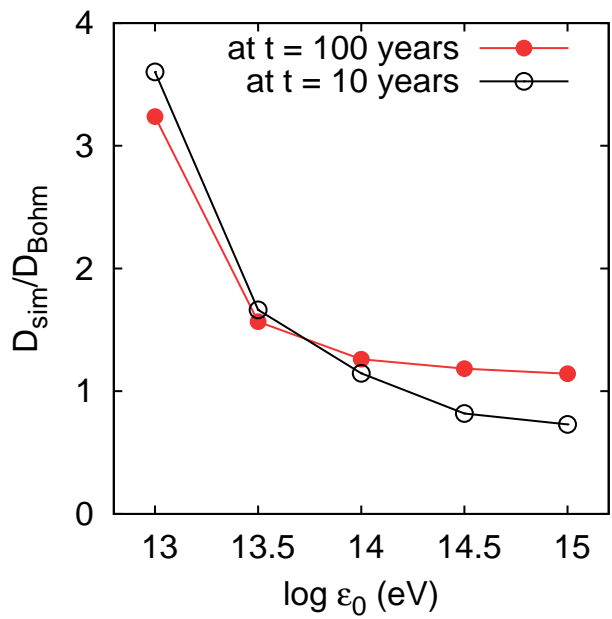}
\caption{The diffusion coefficient as a function
of cosmic ray energy ($\varepsilon_0$).
Vertical axis shows the ratio between the simulation values ($D_{sim}$)
and the estimated values ($D_{Bohm}$) from Equations (\ref{eq:Dbohm}) and (\ref{eq:Dsim}),
shown as a function of $\log \varepsilon_0$ by solid lines.
The results at 10 years and 10${^2}$ years are shown by open black and filled red circles, respectively.}
\label{fig:multi-Bohm}
\end{figure}

Figure \ref{fig:multi-Bohm} shows the ratio of $D_{sim}$/$D_{Bohm}$ 
for different values of $\varepsilon_0$ calculated with Equations (\ref{eq:Dbohm}) and (\ref{eq:Dsim}). 
The values are calculated for 10$^3$ particles at five different values 
of $\varepsilon_0$ at intervals of $\log \varepsilon_0$ = 0.5 eV. 
Black open and red filled circles correspond to $t$ = 10 years and 10${^2}$ years, respectively. 
Both diffusion coefficients, $D_{sim}$ and $D_{Bohm}$, are summarized in Table \ref{table:Dsim/Dbohm-multi}. The value of $D_{sim}/D_{Bohm}$ decreases with increasing $\varepsilon_{0}$ and remains of order unity, 13.5 eV < log $\varepsilon_0$ < 14.5 eV owing to Bohm diffusion process in our simulation.  
Thus, the Bohm diffusion coefficient description for CR diffusion can be justified by 
our test particle simulations, at least for the shocked SNRs.

\subsubsection{Effect of Electric Field}
\begin{figure}[h]
\centering
\includegraphics[scale=1]{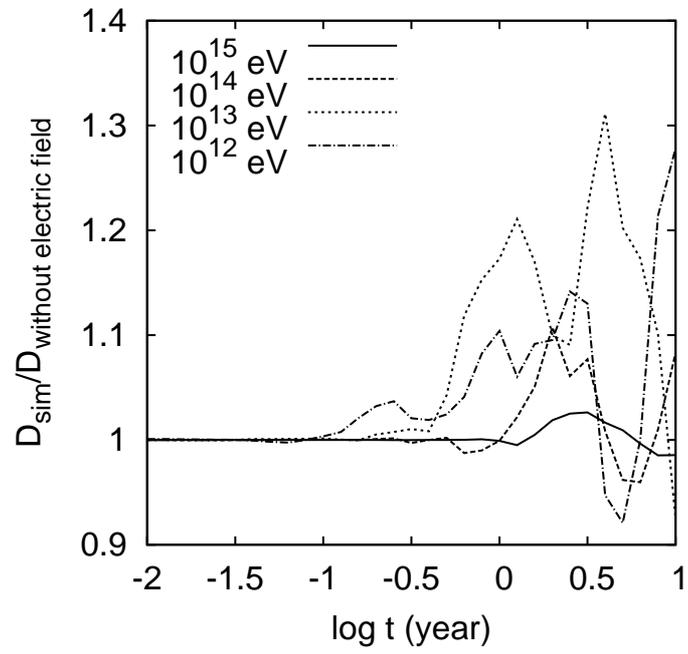}
\caption{Comparison between ratios of results with ${\textit{\textbf{E}}}$ and without ${\textit{\textbf{E}}}$ for the cosmic ray energy range ($\varepsilon_0$) 10 TeV to $10^3$ TeV.}
\label{fig:multi-noE}
\end{figure}

As we mentioned in Section 2.2, the electric field in the comoving frame of the ISM vanishes in the ideal MHD limit. The strength of the electric field in the computation frame is smaller than the magnetic field strength by a factor of $v/c$ (see Equation (\ref{eq:E})). Therefore, the force due to the electric field is small in comparison to the force due to magnetic field. In this Section, we study the effect of electric field on the diffusion of CRs in the ISM. Figure \ref{fig:multi-noE} shows the results of diffusion calculations with and without the background plasma ($\textbf{\textit{E}}$). It shows that inclusion of $\textbf{\textit{E}}$ results in slightly larger diffusion but the difference is rather limited ($<$ 34$\%$). Therefore, electric fields are less important for studying diffusion in configuration space, but they are critical to studying the diffusion in energy space.

\subsection{Diffusion Process in a One-phase Medium}
From this Section, we describe the transport properties of particles in configuration space for a one-phase medium as a function of $\log \it t$, and analyze the diffusion in 
configuration space. 
The values are averaged over $N_{ptls}$ = 10$^3$ particles for each set of 
energies ($\varepsilon_0$). We calculated up to $t$ = 10$^2$ years. 
We mainly demonstrate the particle distribution and background information 
of region-I as given in Table \ref{table:physical-one}. 
This region is chosen to be in the dense, strong magnetic field post-shock region.
The calculation method and conditions for a one-phase medium are the same as a multiphase medium of Section 3.1.

\subsubsection{Diffusion in Configuration Space}
\begin{figure}[h]
\centering
\includegraphics[scale=1]{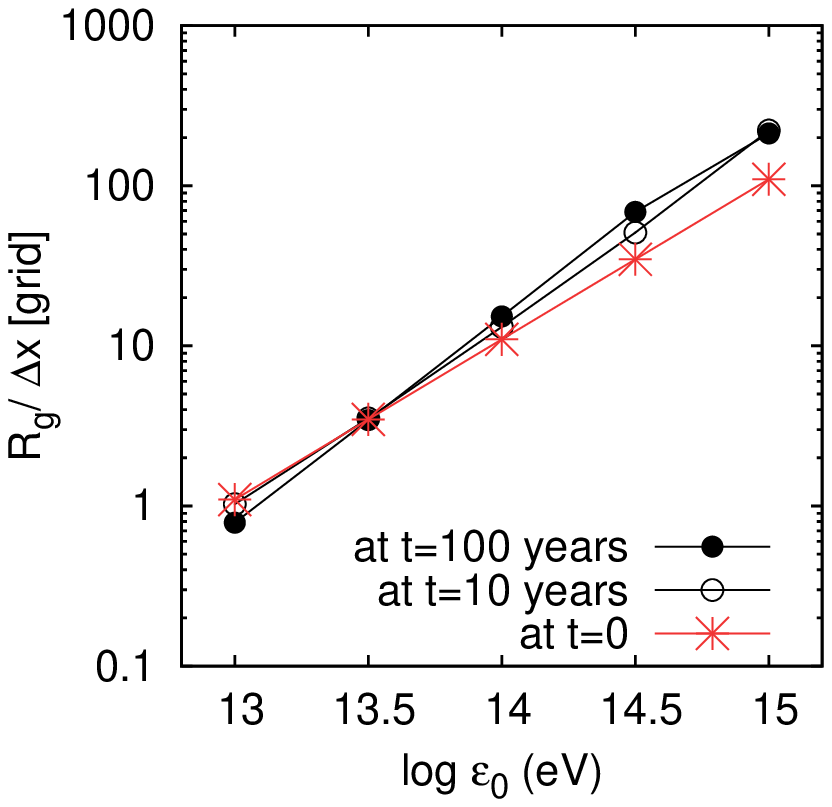}
\caption{The gyroradius ($R_g$) in units of pc as a function of cosmic ray initial energy ($\varepsilon_0$). 
The results at 10 years and 10${^2}$ years are shown by black open circles 
and filled circles, respectively. Red star symbol is for the case $t$ = 0.}
\label{fig:onephase-Rg}
\end{figure}

\begin{figure}[h]
\centering
\includegraphics[scale=1]{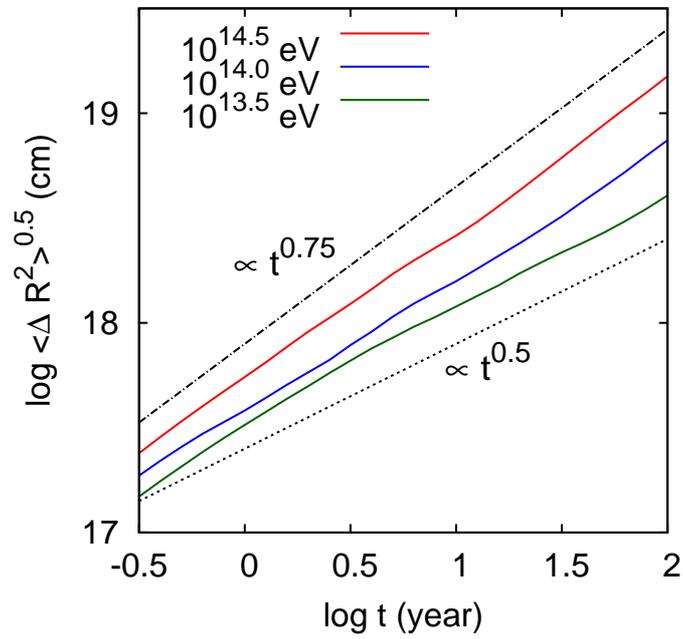}
\caption{Travel distance ($\sqrt{\langle(\Delta R)^2\rangle}$ ) of cosmic ray particles ($\varepsilon_0$) as a function of $\log \textit{t}$. 
Green, blue, and red lines indicate the curves of  
$\varepsilon_0$ = $10^{13.5}$ eV, $10^{14}$ eV and $10^{14.5}$ eV, respectively.
The red and blue lines are similar to the uppermost black dashed line that is proportional to  t$^{0.75}$, corresponding to the fast process (superdiffusive). 
And green line is similar to the uppermost black dashed line that is proportional to t$^{0.5}$, corresponding to the slowest process.} 
\label{fig:onephase-mainresult-rms-t}
\end{figure}

The lines in Figure \ref{fig:onephase-mainresult-rms-t} represent $\sqrt{\langle(\Delta R)^2\rangle}$ 
for protons with energies of $\varepsilon_0$ = $10^{13.5}$ eV, $10^{14}$ eV, 
and $10^{14.5}$ eV, corresponding to $R_g$ $>$ $\Delta x$ and $L_{min} < R_{g} < L_{max}$ (see the gyroradius in Figure \ref{fig:onephase-Rg} and \citep{Inoue2013}). 
The uppermost black dashed line in the figure represents the slope $t^{0.75}$. 
We observed that the evolution of $\sqrt{\langle(\Delta R)^2\rangle}$ for $\varepsilon_0=10^{14}$ eV (blue solid line) and 10$^{14.5}$ eV (red solid line) can be fitted by $t^{0.75}$, corresponding to 
fast diffusion, so-called ``superdiffusion". 
Superdiffusion has been studied by S. Xu $\&$ H. Yan (2013) \citep{XuYan2013} and 
A. Lazarian $\&$ H. Yan (2014) \citep{Lazarian2014}. 
One of the reasons for superdiffusion can be the wandering of 
magnetic field lines in a turbulent medium (see A. Lazarian $\&$ H. Yan (2014) for details). 
We demonstrate about superdiffusion property in next Section 3.2.3. 
In Figure 8, $\sqrt{\langle(\Delta R)^2\rangle}$ for $\varepsilon_0$ = $10^{13.5}$ eV shows standard diffusion,  corresponding to a simple random walk, represented by black dotted line in Figure \ref{fig:onephase-mainresult-rms-t}. 
This random walk can be characterized by a diffusion coefficient, 
as we will discuss in the Section 3.1. 

\subsubsection{Comparison with Bohm Diffusion Rate}
\begin{figure}[h]
\centering
\includegraphics[scale=1]{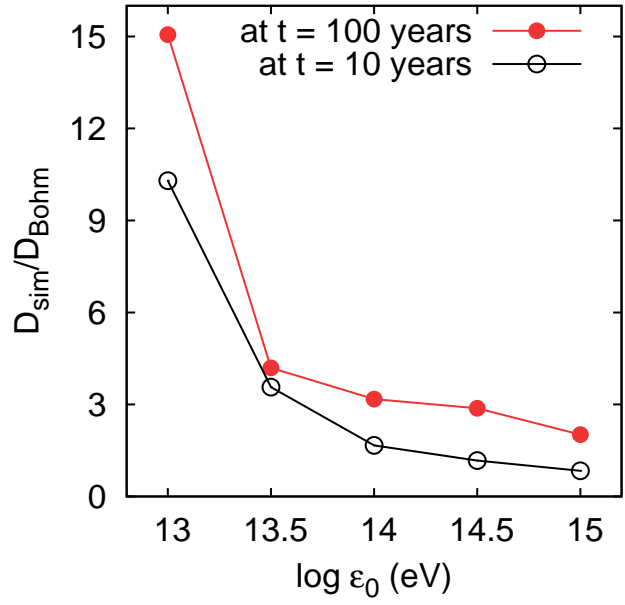}
\caption{The gyroradius and the diffusion coefficient as a function 
of cosmic ray energy. 
The left panel is gyroradius in units of pc, shown by dashed lines. 
The right panel is the ratio between the simulation values ($D_{sim}$) 
and the estimated values ($D_{Bohm}$) from Equations (\ref{eq:Dbohm}) and (\ref{eq:Dsim}), 
shown as a function of $\log \varepsilon_0$ by solid lines. 
The results at 10 years and 10${^2}$ years are shown by open black 
and filled black circles, respectively. Stars are for the case $t$=0.}
\label{fig:onephase-DBohm}
\end{figure}

Figure \ref{fig:onephase-DBohm} shows the ratio of $D_{sim}$/$D_{Bohm}$ for different energies   ($\varepsilon_0$), calculated using Equations (\ref{eq:Dbohm}) and (\ref{eq:Dsim}) (see Section 3.1 for review).  
Applying the Bohm diffusion process, the differences in $D_{sim}$/$D_{Bohm}$ are less than a factor of 5 with energy range with 13.5 eV < log $\varepsilon_0$ < 14.5 eV, at both times in Figure \ref{fig:onephase-DBohm}. 
The differences in the ratio of $D_{sim}$/$D_{Bohm}$ between $t$ = 10 years and $10^{2}$ years have grown more than those in Figure \ref{fig:multi-Rg} due to superdiffusive process.
However, because the differences in $D_{sim}$/$D_{Bohm}$ are significantly small, thus our result precisely justifies the Bohm diffusion. 
The diffusion coefficients for a one-phase medium are summarized in Table \ref{table:one-phase-diffusioncoeff}.

\subsubsection{Superdiffusion}
Superdiffusion has been reported in simulations by \citep{Lazarian2014, XuYan2013} that is  expected to occur below the injection scale. (see the injection scale of one-phase medium case in Figure \ref{fig:FourierPowerSpectrum}). 
We compare our results with S. Xu \& H. Yan (2013) to investigate the 
existence of superdiffusion in our simulations. 
The pre-shock magnetic field lines in our simulations are laminar 
and not turbulent, and the turbulence is created by the RMI \citep{Inoue2012} 
driven by the propagation of a blast wave. 
In contrast, S. Xu \& H. Yan (2013) drove turbulence by solenoidal forcing in Fourier space. 

\begin{figure}[h]
\centering
\includegraphics[scale=1]{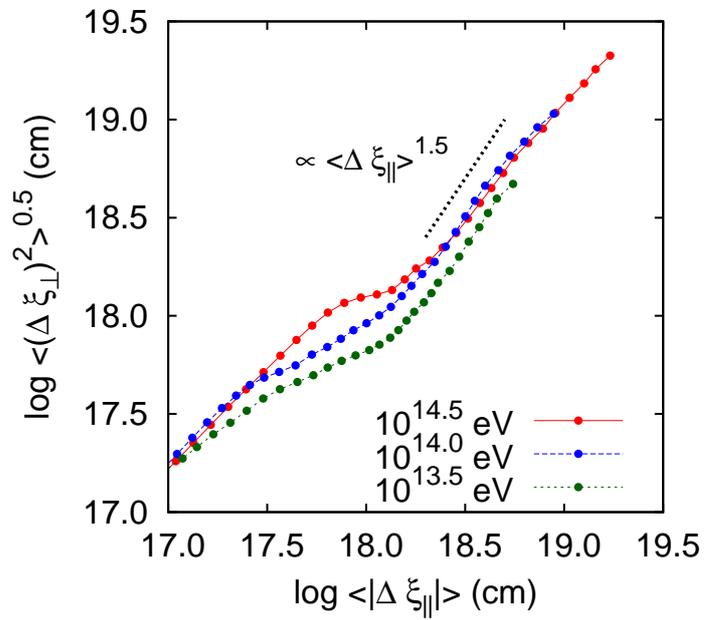}
\caption{Relation between travel distances in the directions parallel and perpendicular 
to the magnetic field lines. 
Green, blue, and red lines indicate the curves of  
$\varepsilon_0$ = $10^{13.5}$ eV, $10^{14}$ eV and $10^{14.5}$ eV, respectively.
The black dotted line describes $\langle(\Delta \xi_{\perp})^2\rangle^{0.5}$ $\propto$ $\langle \vert\Delta \xi_{\parallel} \vert\rangle^{1.5}$ in limited time.}
\label{fig:xu}
\end{figure}

We calculated the dispersion $\sqrt{\langle(\Delta \xi_{\perp})^2}$ 
as a function of $\langle \vert\Delta \xi_{\parallel} \vert\rangle$ up to 10${^2}$ years, 
the result of which is shown in Figure \ref{fig:xu}, where $\xi_{\perp}$ is perpendicular to 
the magnetic field (\textbf{\textit{B}}), and $\xi_{\parallel}$ is 
the displacement along the magnetic field. 
The upper black dashed lines represent 
$\sqrt{\langle(\Delta \xi_{\perp})^2}$ $\propto$ $t^{0.75}$, respectively.

The slope close to 1.5 may a manifestation of wandering magnetic field lines analogous 
to Richardson diffusion as discussed in A. Lazarian \& H. Yan 2013. 
This may the reason for the superdiffusion found in Figure \ref{fig:onephase-mainresult-rms-t}. 
However, the larger energy particles do not show the 1.5 slope of 
Figure \ref{fig:xu}, because the gyroradius of larger energy particles is 
larger than the injection scale of turbulence.

\section{Summary and Discussion}
Young supernova remnants are accompanied by collisionless shock waves in which DSA 
is expected to occur. Gamma rays are emitted by pion decay caused by CR protons 
accelerated in SNRs interacting with molecular clouds. 
Observations of gamma rays from the vicinity of SNRs have shown strong evidence 
that galactic CR protons are accelerated by the shock waves of the SNRs \citep{Ackermann2013}. 
In DSA theory, efficient acceleration of CRs requires a small diffusion coefficient. 
Indeed, the time-variability seen in synchrotron X-ray observations of 
{\em RX J1713.7-3946} seems to need the slowest diffusion process \citep{Uchiyama2007}. 
T. Inoue et al. \citep{Inoue2012} adopted Bohm diffusion as a diffusion of CRs in the same SNR to explain both the gamma-ray spectrum and X-ray variability with magnetic field amplification up to mG.
In previous research, Bohm diffusion is discussed by considering 
($B^{2}/\delta B^2$ $\sim$ 1) (see. Equation (\ref{eq:skilling}), \citep{Inoue2012}, and \citep{Uchiyama2007}).

In the present study, we performed test particle simulations that describe 
the propagation of high-energy CR particles in SNRs in the early evolutionary phase. 
%We investigate the diffusion of CR particles in a background medium that is created from numerical simulation of a realistic interstellar medium, and find it is well described by Bohm diffusion.
We investigate the diffusion of CR particles in a background medium that is created from numerical simulation of a realistic interstellar medium. %, and find it is well described by Bohm diffusion.
Our results show that the Richtmeyer-Meshkov instability can provide enough turbulence downstream of the shock to make the diffusion coefficient close to the Bohm level for energy larger than 30 TeV.
Thus, the present study indicates that we can use Bohm diffusion coefficient
in practical calculations of CR particle diffusion, which may support
the assumption used in the interpretation of high-energy emission from SNRs \citep{Inoue2012}.
 
It is worth stressing that in this paper we do not deal with the diffusion of particles in the medium upstream of the shock. We recall that the DSA mechanism can predict the maximum energy of particles needed to explain the observations, only if the diffusion coefficient is close to the Bohm one also in the upstream.  
 
We also identified superdiffusion in a one-phase medium, in which the travel distance is proportional to $t^{0.75}$. 
Because the period of superdiffusion is limited in time, the actual travel distance is not so different from that described by a Bohm diffusion coefficient, at least for particles with energies above 30 TeV. 
 
\clearpage

\section*{Acknowledgements}
We thank Toshio Terasawa, Takeru K. Suzuki, and Jennifer Stone for useful comments. 
SR thanks Kazunari Iwasaki and Takuma Matsumoto for their kind assistance. 
Numerical computations were in part carried out on Cray XC30 at Center for Computational Astrophysics, National Astronomical Observatory of Japan.

{}

\begin{table}[h]
\centering
\begin{tabular}{ccccc}
\hline
 & region-I & post-shock region &  \\ \hline \hline
$\langle\textbf{B}\rangle$ (G)& 2.36e-5 & 2.13e-5     \\
$\sqrt{\langle\delta \textbf{B}\rangle^2}$ (G)& 3.36e-5  & 3.53e-5   \\
$\langle\rho\rangle$ (cm$^{-3}$) & 2.15 & 5.65  \\
$\sqrt{\langle\delta \rho\rangle^2}$ (cm$^{-3}$) & 1.84 &  5.22 \\
\end{tabular}
\caption{Physical values in different simulation regions for the multiphase medium. ``region-I'' and
``post-shock region" indicate the volume sizes occupied by the initial positions
of particles in our simulation. region-I and post-shock region have volumes (0.3 pc)$^{3}$ for the range 1.5 - 1.8 pc in $xyz$ configuration space and (2 pc)$^{3}$ in 1.6 - 1.8 pc of $x$, 0.7 - 0.9 pc of $y$, and 1.4 - 1.6 pc of $z$ configuration space. In this paper, we show the results using region-I for a multiphase medium.}
\label{table:physical-multi}
\end{table}

\begin{table}[h]
\centering
\begin{tabular}{ccc|cc}
& \multicolumn{2}{c}{t = 10 years} & \multicolumn{2}{c}{t = 100 years}                                                         \\ \hline
$\log \varepsilon_{0}$ (eV) & D$_{sim}$ (cm$^{2}$/s) & D$_{Bohm}$ (cm$^{2}$/s) & \multicolumn{1}{c}{D$_{sim}$ (cm$^{2}$/s)} & \multicolumn{1}{c}{D$_{Bohm}$ (cm$^{2}$/s)} \\ \hline  \hline
15.0  & 5.839e+27   & 8.000e+27    & 7.520e+27   & 6.584e+27                                     \\
14.5  & 2.044e+27   & 2.497e+27    & 2.560e+27    & 2.163e+27
\\
14.0  & 9.054e+26  & 7.906e+26    & 8.116e+26   & 6.437e+26                                     \\
13.5  & 4.745e+26  & 3.193e+26     & 3.193e+26   & 2.038e+26
\\
13.0  & 3.314e+26  & 2.015e+26     & 2.015e+26   & 6.225e+25
\end{tabular}
\caption{Differences between $D_{sim}$ vs. $D_{Bohm}$ as a function of $\varepsilon_0$. We used the region-I volume results shown in Table \ref{table:physical-multi}.}
\label{table:Dsim/Dbohm-multi}
\end{table}

\begin{table}[h]
\centering
\begin{tabular}{ccccc}
\hline
 & region-I & region-II & post-shock region &  \\ \hline \hline
$\langle\textbf{B}\rangle$ (G)& 1.262e-5 & 6.033e-6 & 1.102e-5     \\
$\sqrt{\langle\delta \textbf{B}\rangle^2}$ (G)& 1.092e-5 & 9.306e-6 & 1.123e-5   \\
$\langle\rho\rangle$ (cm$^{-3}$) & 1.303 & 0.12 & 1.025  \\
$\sqrt{\langle\delta \rho\rangle^2}$ (cm$^{-3}$) & 1.017 & 0.313 &  0.917 \\
\end{tabular}
\caption{Physical values in different simulation regions for one-phase medium. ``region-I'', ``region-II'', and 
``post-shock region" indicate the volume sizes occupied by the initial positions 
of particles in our simulation. region-I and region-II have volumes of (0.3 pc)$^{3}$ for the range 0.7 - 1 pc in $xyz$ configuration space of our simulated data and 
(0.2 pc)$^{3}$ for 0.8 - 1.0 pc in $x$, 0.7 - 0.9 pc in $y$, and 1.0 - 1.2 pc in $z$  configuration space of our simulated data. These regions are chosen to be in the dense, strong magnetic field post-shock region. Our simulation is performed in region-I for a one-phase medium.}
\label{table:physical-one}
\end{table}

\begin{table}[h]
\centering
\begin{tabular}{ccc|cc}
& \multicolumn{2}{c}{t = 10 years} & \multicolumn{2}{c}{t = 100 years}                                                         \\ \hline
$\log \varepsilon_{0}$ (eV) & D$_{sim}$ (cm$^{2}$/s) & D$_{Bohm}$ (cm$^{2}$/s) & \multicolumn{1}{c}{D$_{sim}$ (cm$^{2}$/s)} & \multicolumn{1}{c}{D$_{Bohm}$ (cm$^{2}$/s)} \\ \hline  \hline
15.0  & 1.120e+28   & 1.334e+28    & 2.575e+28   & 1.277e+28                                     \\
14.5  & 3.593e+27   & 3.073e+27    & 1.189e+28   & 4.134e+27                                     \\
14.0  & 1.318e+27   & 7.920e+26    & 2.916e+27   & 9.189e+26                                     \\
13.5  & 7.578e+26   & 2.126e+26    & 8.687e+26   & 2.071e+26                                     \\ 
13.0  & 6.374e+26   & 6.191e+25    & 7.136e+26   & 4.740e+25                                     
\end{tabular}
\caption{Difference between $D_{sim}$ and $D_{Bohm}$ as a function of $\varepsilon_0$.}
\label{table:one-phase-diffusioncoeff}
\end{table}

\end{document}